\documentclass[aps,pra,twocolumn,superscriptaddress,showpacs,showkeys,amsmath,amssymb]{revtex4}

\usepackage{amsfonts}
\usepackage{amssymb,amsmath}
\usepackage{mathrsfs}
\usepackage{latexsym}
\usepackage{amsmath}
\usepackage[cp1251]{inputenc}
\usepackage{graphicx}
\usepackage{dcolumn}
\usepackage{bm}
\usepackage{color}

\begin{document}

    \title{Mean-field properties of impurity in Bose gas with three-body forces}

    \author{Volodymyr~Pastukhov\footnote{e-mail: volodyapastukhov@gmail.com}}
    \affiliation{Department for Theoretical Physics, Ivan Franko National University of Lviv,\\ 12 Drahomanov Street, Lviv-5, 79005, Ukraine}

    \date{\today}



    \begin{abstract}
We exactly analyze, on the mean-field level, the low-momentum properties of a single impurity atom loaded in the dilute one-dimensional Bose gas with two- and three-body short-range interactions. Particularly the Bose polaron binding energy and the quasiparticle residue are calculated for the considered system in the broad region of parameters change. We also explore the generic mean-field formula for the polaron effective mass which was shown to depend on the density profile of bath particles with a motionless impurity immersed.
    \end{abstract}

    \maketitle

\section{Introduction}
\label{sec1}
\setcounter{equation}{0}
Talking about impurities in condensed matter systems we mainly keep in mind the situation when a very small number of atoms is immersed in the majority of bath particles. Then in the linear approximation over the density of extraneous particles both an effects of the impurity statistics and the impurity-impurity interaction can be freely neglected and one faces the problem of a single particle loaded in the many-body medium. When this medium is formed by cold Bose-condensed atoms the problem is usually called the Bose polaron one. In the past few years these low-temperature objects, which are very interesting from theoretical point of view (see for review \cite{Grusdt_Demler}) and can be realized in the ultracold-atoms setup \cite{Schmid_et_al,Spethmann_et_al}, attracted much attention of experimentalists \cite{Hu,Jorgenzen}, which in turn, stimulated numerous theoretical studies \cite{Rath,Shashi,Li,Christensen,Grusdt_15,Volosniev_15,Vlietinck_et_al,Ardila_1,Shchadilova,Ardila_2,Kain_Ling_16,Grusdt_17,Panochko_17,Lampo,Grusdt_18,Pastukhov_CrBP,Panochko_18,Mehboudi}. 

The most promising candidates for the observation of non-trivial quantum phases and for the potential application in engineering of quantum devices \cite{Lychkovskiy_18}, however, are the low-dimensional Bose polarons. Particularly the dynamics \cite{Burovski,Gamayun} and thermodynamical properties of an impurity interacting via a short-range potential with the surrounding one-dimensional Bose bath with a contact interaction between particles was extensively studied experimentally \cite{Catani_et_al}, by means of quantum \cite{Parisi} and diffusion \cite{Grusdt} Monte Carlo simulations or using various approximate techniques \cite{Ovchinnikov,Dehkharghani,Volosniev,Pastukhov_1D,Kain_Ling_18,Mistakidis}. Moreover, in the equal-mass limit this model is exactly solvable \cite{McGuire_65,McGuire_66}. However, the Bethe ansatz wave function underlying this solution can be applied to Bose particles with a pairwise contact potential only, and the inclusion of higher-order few-body interactions excludes the exact consideration. On the other hand, very recently the properties of one-dimensional Bose gas with a three-body interactions \cite{Sekino,Pastukhov_3BI} has become the subject of lively discussions \cite{Pricoupenko,Guijarro,Nishida}. The Bose polaron problem, to our knowledge, was not previously considered in such media, therefore, the objective of present article is to explore the properties a moving impurity in a dilute 1D bath formed by Bose particles interacting via the two- and three-body short-range potentials.
 
\section{Formulation}
So in the following we discuss the motion of a single impurity atom immersed in the one-dimensional bath formed by $N$ Bose particles. The system is assumed to be loaded in a large volume $L$ (the periodic boundary conditions are imposed) and our discussion is limited to the examination of thermodynamic limit ($N\to \infty$, $L\to \infty$ but $N/L=\bar{n}$). Adopting the second-quantized description for bosons the appropriate Hamiltonian reads
\begin{eqnarray}\label{H}
H=-\frac{\hbar^2}{2m_I}\partial^2_{z_I}+H_{\rm int}+H_B,
\end{eqnarray}
where the first term is the kinetic energy of impurity which position is denoted by $z_I \in [-L/2, L/2]$, and the second one
\begin{eqnarray}\label{H_int}
H_{\rm int}=g\psi^{+}(z_I)\psi(z_I),
\end{eqnarray}
that describes the delta-like two-body interaction between impurity and bath-forming particles which strength is characterized by coupling constants $g$. The Hamiltonian of Bose particles with mass $m$ and chemical potential $\mu$
\begin{eqnarray}\label{H_B}
H_B=\int dz\, \psi^+(z)\left\{-\frac{\hbar^2}{2m}\partial^2_z
-\mu\right\}\psi(z)\nonumber\\
+\sum_{l=2,3}\frac{g_l}{l!}\int dz [\psi^+(z)]^l[\psi(z)]^l,
\end{eqnarray}
includes two- and three-body short-range potentials with coupling parameters $g_2$ and $g_3$, respectively. The field operators $\psi^{+}(z)$ and $\psi(z)$ presented in (\ref{H_B}) satisfy the standard bosonic commutation relations $[\psi(z), \psi^+(z')]=\delta(z-z')$, $[\psi(z), \psi(z')]=0$. The further consideration is typical for the Bose polaron problem: taking into account the conservation of total momentum we perform the Lee-Low-Pines \cite{LLP} unitary transformation $H'=U^{+}HU$ with generator $U=\exp\left\{iz_I(p-\mathcal{P})\right\}$, where $\hbar p$ should be treated as the impurity momentum and operator $\hbar\mathcal{P}=-i\int dz\, \psi^+(z)\hbar\partial_z\psi(z)$ denotes the momentum of Bose particles. The unitary-transformed Hamiltonian $H'$ commutes with the impurity momentum operator which eigenvalue can be chosen arbitrary ($\hbar p$ in our case) and contains extra term
\begin{eqnarray}\label{H_prime}
H'=\frac{\hbar^2p^2}{2m_I}+H'_{\rm int}+H_B+\Delta H',
\end{eqnarray}
where the second term now reads $H'_{\rm int}=g\psi^{+}(0)\psi(0)$, while the last one $\Delta H'$ represents both the effective external potential and induced two-body interaction between bosons associated with the impurity motion
\begin{align}\label{DH}  
 \Delta H'=-\frac{\hbar^2p}{m_I}\mathcal{P} +\frac{\hbar^2}{2m_I}\mathcal{P}^2.
\end{align}
In general, the model (\ref{H_prime}) has no exact solutions. Some limiting cases $g_3=0$ and $m_I=m$ (or $m_I\gg m$) can be studied by means of the Bethe ansatz wave function and are well-discussed in the literature. The inclusion of the three-body forces, however, completely breaks the integrability of the model and therefore our further analysis requires some approximations. In the present article for the formulated problem we adopt the mean-field treatment \cite{Astrakharchik_04} that goes back to the work of Gross \cite{Gross} and was recently shown \cite{Volosniev} to allow the analytical description of 1D Bose polarons. The main idea is to replace in the normal-ordered Hamiltonian $H'$ the Bose field operators by some complex function $\psi^+(z)\to \psi^*(z)$ and $\psi(z)\to \psi(z)$, which minimizes the appropriate energy functional. This replacement is identical to the use of the trial ground-state wave function $\Psi\propto \prod_z|\psi(z)\rangle $ written in coherent states representation or $\Psi\propto \prod_{1\le j \le N}\psi(z_j-z_I)$ explicitly in the coordinate representation for the initial Hamiltonian (\ref{H}). Of course, such a simplified consideration that is only applicable to Bose particles and substantially facilitates the theoretical description of the problem is not without flaws. Particularly, it fully neglects important quantum correlations which are responsible for the emergence of the off-diagonal quasi-long-range order in the 1D Bose systems at zero temperature. On the other side, the mean field approximation is known to give the correct first-order thermodynamics of weakly-interacting Bose systems, which narrows the range of applicability of the obtained results to the case of dilute Bose environments.
 
Before we explicitly write down the equation for the unknown function that extremizes energy functional it is convenient, in context of further calculations especially at non-zero $\hbar p$, to introduce the Madelung parametrization $\psi^*(z)=\sqrt{n(z)}e^{-i\varphi(z)}$, $\psi(z)=\sqrt{n(z)}e^{i\varphi(z)}$, where the amplitude $n(z)$ is taken in our case as a microscopic density of bosons at position $z$. Now, the couple of the Euler-Lagrange equations that determine the phase $\varphi(z)$ and density $n(z)$ fields reads
\begin{align}\label{E_L_1}  
-\frac{\hbar^2}{4m_r}\frac{n_{zz}}{n}+\frac{\hbar^2}{8m_r}\frac{n_{z}^2}{n^2}+\frac{\hbar^2}{2m_r}\varphi_z^2 +g_2n+\frac{1}{2}g_3n^2\nonumber\\
+g\delta(z)-\mu-\frac{\hbar^2}{m_I}\varphi_z(p-\mathcal{P})=0,
\end{align}
\begin{align}\label{E_L_2}  
-\frac{\hbar^2}{m_r}\partial_z(n\varphi_{z})+\frac{\hbar^2}{m_I}(p-\mathcal{P})n_z=0,
\end{align}
where the subscript $z$ denotes the differentiation with respect to position. The impurity-boson reduced mass $m_r=mm_I/(m+m_I)$ occurred in the above formulae due to normal ordering of the last term in Eq.~(\ref{DH}). We have also used the same notation $\hbar\mathcal{P}=\int dz\, n\hbar\varphi_z$ for the expectation value of the Bose gas total momentum calculated in the mean-field approximation.

The search for the solution of the system of coupled equations (\ref{E_L_1}), (\ref{E_L_2}) should be started from solving the second one. The formal solution, up to a constant which provides periodic boundary conditions but is of order $1/L$ and thus is irrelevant for our calculations, yields
\begin{eqnarray}\label{varphi_z}
	\varphi_z(z)=\frac{m_r}{m_I}(p-\mathcal{P})[1-n_{\infty}/n(z)],
\end{eqnarray}
where $n_{\infty}=n(\pm L/2)$. Note that expression (\ref{varphi_z}) ensures that the value of $\mathcal{P}$ is of order unity. Then the formula for $\varphi_z$ should be substituted in the first equation of the system to find the density profile $n(z)$ of Bose particles with the immersed impurity moving with momentum $\hbar p$. Actually, this equation for a given $p$ should be solved self-consistently with the mean-field value of quantity
\begin{eqnarray}\label{P}
\mathcal{P}=\frac{p\Delta_p}{1+\Delta_p}, \ \ 
\Delta_p=\frac{m_r}{m_I}\int dz\,\frac{(n-n_{\infty})^2}{n},
\end{eqnarray}
where $\Delta_p$ also depends on $p$. The above form of $\mathcal{P}$ was obtained from the second Euler-Lagrange equation (\ref{E_L_2}) by multiplying the both sides on $\varphi$ with the subsequent integration by parts. The ground-state energy of system ``Bose gas + moving impurity'' then reads
\begin{eqnarray}\label{E}
E_p=\mu N-\frac{g_2}{2}\int dz\, n^2-\frac{g_3}{3}\int dz\, n^3\nonumber\\
+\frac{\hbar^2}{2m_I}(p^2-\mathcal{P}^2).
\end{eqnarray}
The energy of the polaron $\varepsilon_p=E_p-E_0|_{g=0}$ can be easily calculated as a difference of (\ref{E}) and the mean-field energy of a homogeneous Bose system without impurity.

An important quantity that determines the polaron density of states near its peak, in measurements when a small amount of impurity atoms is loaded in Bose condensate, is the quasiparticle residue $Z_p$. In the presented techniques it is given by the modular square of the overlap integral $Z_p=|\langle \Psi_0|\Psi\rangle|^2$ between the ground state of the homogeneous Bose system $\Psi_0$ and the wave function in the presence of an impurity. For one-dimensional Bose polarons at absolute zero \cite{Grusdt,Pastukhov_1D} and for two-dimensional ones at finite temperatures \cite{Pastukhov_2DBP} this quantity demonstrates a singular behavior (known as an orthogonality catastrophe in 1D). The mean-field ansatz, however, does not capture these peculiarities in 1D giving a finite $p$-dependent result 
\begin{eqnarray}\label{Z}
\ln Z_p=-\bar{n}\int dz\, \left[n/n_{\infty}+1\right.\nonumber\\
\left.-2\sqrt{n/n_{\infty}}\cos(\varphi-\varphi_{\infty})\right],
\end{eqnarray}
(where $\varphi_{\infty}=\varphi(\pm L/2)$) in the large-$N$ limit.

\section{Low-momentum polaron spectrum}
Unfortunately, we did not succeed in obtaining the full analytic solution of the presented problem at arbitrary $p$, i.e., we did not find the energy of a moving polaron in the most general case. The global features of the low-energy polaronic spectra, however, are known without exact solution of the problem. This dependence is quadratic over the momentum of the moving particle
\begin{eqnarray}  
\varepsilon_{p \to 0}=\varepsilon_0+\frac{\hbar^2p^2}{2m^*_I}+\ldots,
\end{eqnarray}
where the binding energy $\varepsilon_0$ can be calculated using the density profile of the Bose gas with a motionless impurity immersed. The effective mass $m^*_I$ is also affected by the interaction with the medium, and now we are going to argue that it is also determined by the spatial distribution of the bath particles at zero impurity momentum. Suppose that we have found the solution when $p=0$, where the problem simplifies substantially. Indeed, from Eq.~(\ref{E_L_2}) we find $\varphi_z=0$ (and consequently $\mathcal{P}=0$) and the density profile of bosons is given by simple equation, which in our case, allows for the existence of an exact analytic solution in terms of elementary functions. Then the polaron binding energy is easily calculated. Moreover, with this solution in hand and by using Eq.~(\ref{P}), we are in position to obtain the coefficient that determines the leading-order linear dependence of quantity $\hbar\mathcal{P}$ on momentum $\hbar p$ when impurity starts to move slowly. The latter enables the calculations of the correction to the total energy functional in the small-$p$ limit 
\begin{eqnarray}  
\frac{\hbar^2}{2m_r}\int dz \,n\varphi_z^2+\frac{\hbar^2}{2m_I}(p-\mathcal{P})^2\nonumber\\
=\frac{\hbar^2p^2}{2m_I}\frac{1}{1+\Delta_{0}}=\frac{\hbar^2p^2}{2m^*_I},
\end{eqnarray}
which actually determines the Bose polaron effective mass. Finally, it should be noted that the above formula is not restricted to the considered problem and can be easily used to obtain the so-called hydrodynamic effective mass of an impurity for any other mean-field-like treatments. Because the only information required for its evaluation is the density profile of the bath-forming particles in the presence of a static impurity, which in principle, can be approximately calculated by means of the local density approximation for arbitrary media.

Equation (\ref{E_L_1}) at zero $p$ reduces to the following one:
\begin{eqnarray}\label{MF_densprof}
-\frac{\hbar^2}{2m_r\sqrt{n}}\partial^2_{z}\sqrt{n}+g_2n+\frac{1}{2}g_3n^2+g\delta(z)=\mu,
\end{eqnarray}
which generic solution without three-body forces (i.e. when $g_3$=0) is well-discussed in literature \cite{Carr,DAgosta} for an arbitrary number of Bose particles loaded in a finite volume $L$, but in the thermodynamic limit ($L\to \infty$) only phase with $g_2>0$ is stable. 
In a case of non-zero three-body coupling the situation is a little bit more complicated. The chemical potential equals to $\bar{n}g_2+\bar{n}^2g_3/2$ and the uniform phase is thermodynamically stable when $g_3>-g_2/\bar{n}$ both for repulsive ($g_2>0$) and attractive ($g_2<0$) two-body interaction. Below we restrict our analysis to the experimentally relevant case of a dilute Bose bath with the two-body repulsion $g_2>0$ only and additionally we assume that the system without impurity does not collapse (i.e., condition $g_3>-g_2/\bar{n}$ is always satisfied). Imposing periodic boundary conditions $n(\pm L/2)=n_{\infty}$ in the thermodynamic limit ($n_{\infty}={\rm finite}$) we immediately obtain
\begin{eqnarray}\label{MF_sol}
\frac{n(z)}{n_{\infty}}=\frac{\cosh(2\kappa|z|+y_0)-\sigma}{\cosh(2\kappa|z|+y_0)+\sigma\gamma},
\end{eqnarray}
here $\sigma={\rm sign}(g)$, the density at infinity is defined as follows $g_2n_{\infty}+g_3n_{\infty}^2/2=\mu$, the chemical potential is determined by the normalization $\int dz\, n=N$, $\kappa =\sqrt{ m_r(g_2n_{\infty}+g_3n_{\infty}^2)}/\hbar$ and  $\gamma =\frac{\eta_2+4\eta_3/3}{\eta_2+2\eta_3/3}$, where dimensionless couplings equal to $\eta_2=mg_2/(\hbar^2n_{\infty})$ and $\eta_3=mg_3/\hbar^2$, respectively. Parameter $y_0$ is fixed by the boundary condition originating from the delta-function term in Eq.~(\ref{MF_densprof}) and in general is determined by the cubic equation. After the introduction of additional notation $\xi=m_rg/(\hbar^2\kappa)$ the explicit expressions for $y_0$ are
\begin{eqnarray}
 &&\coth(y_0/2)=\xi/3+(2/3)\sqrt{3+\xi^2}\nonumber\\
 &&\times\cos\left\{\frac{1}{3}\arccos\left[ \frac{\xi^3+\frac{9\xi\eta_2}{2(\eta_2+\eta_3)}}{(3+\xi^2)^{3/2}}
\right]\right\},
\end{eqnarray}
for repulsive ($g>0$) and
\begin{eqnarray}
&&\tanh(y_0/2)=-|\xi|/3+(2/3)\sqrt{3+\xi^2}\nonumber\\
&&\times\cos\left\{\frac{\pi}{3}-\frac{1}{3}\arccos\left[\frac{|\xi|^3+\frac{9|\xi|\eta_2}{2(\eta_2+\eta_3)}}{(3+\xi^2)^{3/2}}
\right]\right\},
\end{eqnarray}
for attractive ($g<0$) Bose polarons. Within the density profile (\ref{MF_sol}) all parameters determining the low-energy impurity spectrum, namely, the binding energy and the effective mass can be expressed in terms of elementary functions. The appropriate formulae are too cumbersome and therefore not written.

It is looking natural to start the discussion of the obtained results from the comparison of two limiting case, namely, when the Bose particles interact through the pairwise potential ($g_3=0$ and $g_2>0$), and when the three-body interaction is the only one that is switched on ($g_2=0$ and $g_3>0$). The behavior of the impurity in the intermediate region, where all $g_l$s are positive, can be understood from the combination of these limits. Introducing the energy scale $\hbar^2\bar{n}^2/m$, dimensionless parameter $\eta=mg/(\hbar^2n_{\infty})$ (at this stage of calculations $n_{\infty}$ can be freely replaced by $\bar{n}$) that sets the strength of the boson-impurity interaction, magnitude $w=m/m_I$ of the Bose particles mass in units of the bare mass of the impurity, we presented dependences of the low-energy polaron spectrum graphically in Figs.~1-3.
\begin{figure}[h!]
	\centerline{\includegraphics
		[width=0.45\textwidth,clip,angle=-0]{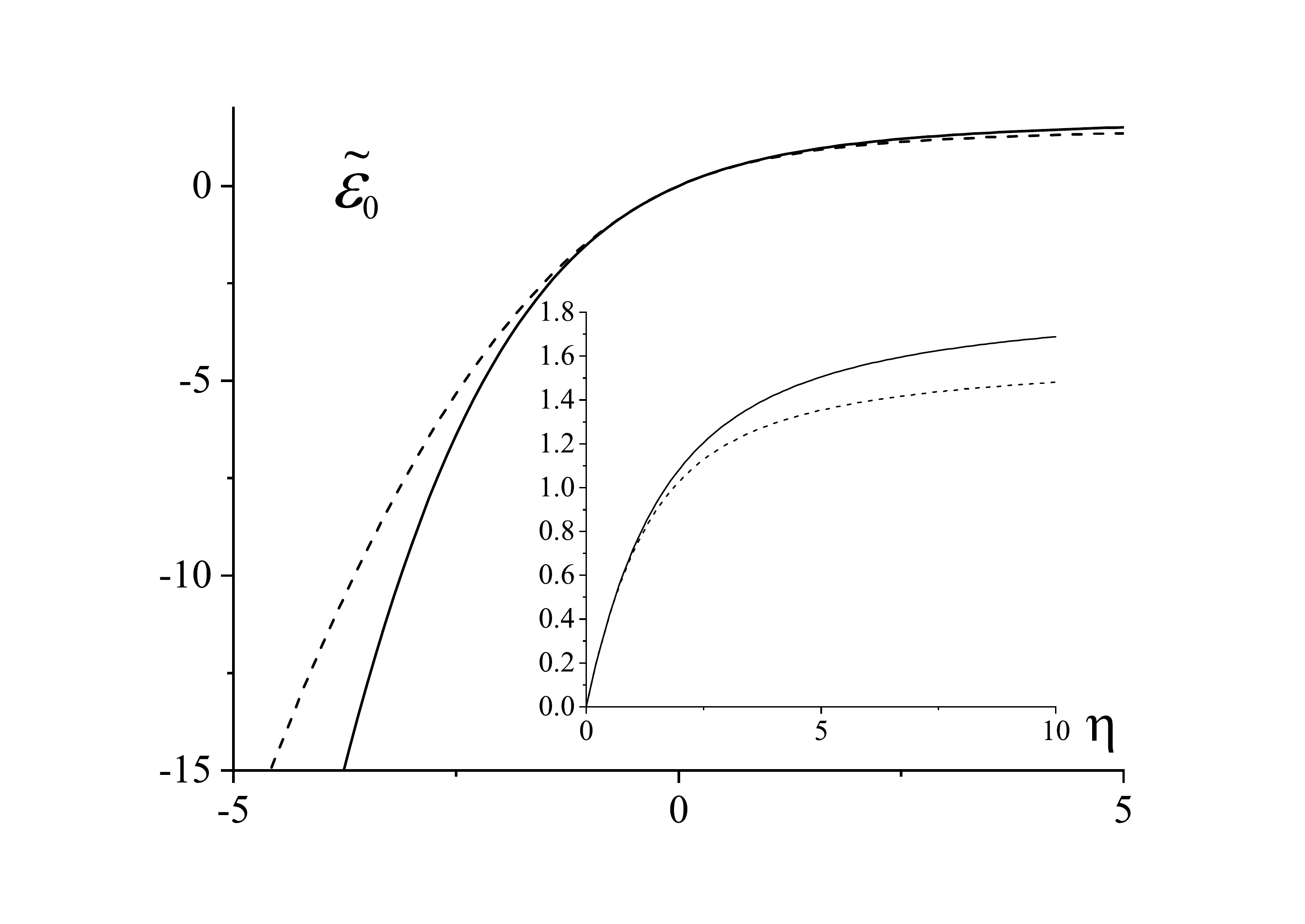}}
	\caption{Graphical dependence of the polaron binding energy $\varepsilon_0$ (in units of $\hbar^2\bar{n}^2/m$) on the boson-impurity coupling constant. Solid and dashed curves correspond to the Bose bath with fully two- and three-body repulsion, respectively. Another parameters are chosen in the following way: mass ratio $w=1$, $\eta_2=1$ (solid line) and $\eta_3=1$ (dashed line). The inset shows two curves at the positive semi-axis, i.e., for the repulsive Bose polaron.}
\end{figure}
\begin{figure}[h!]
	\centerline{\includegraphics
		[width=0.45\textwidth,clip,angle=-0]{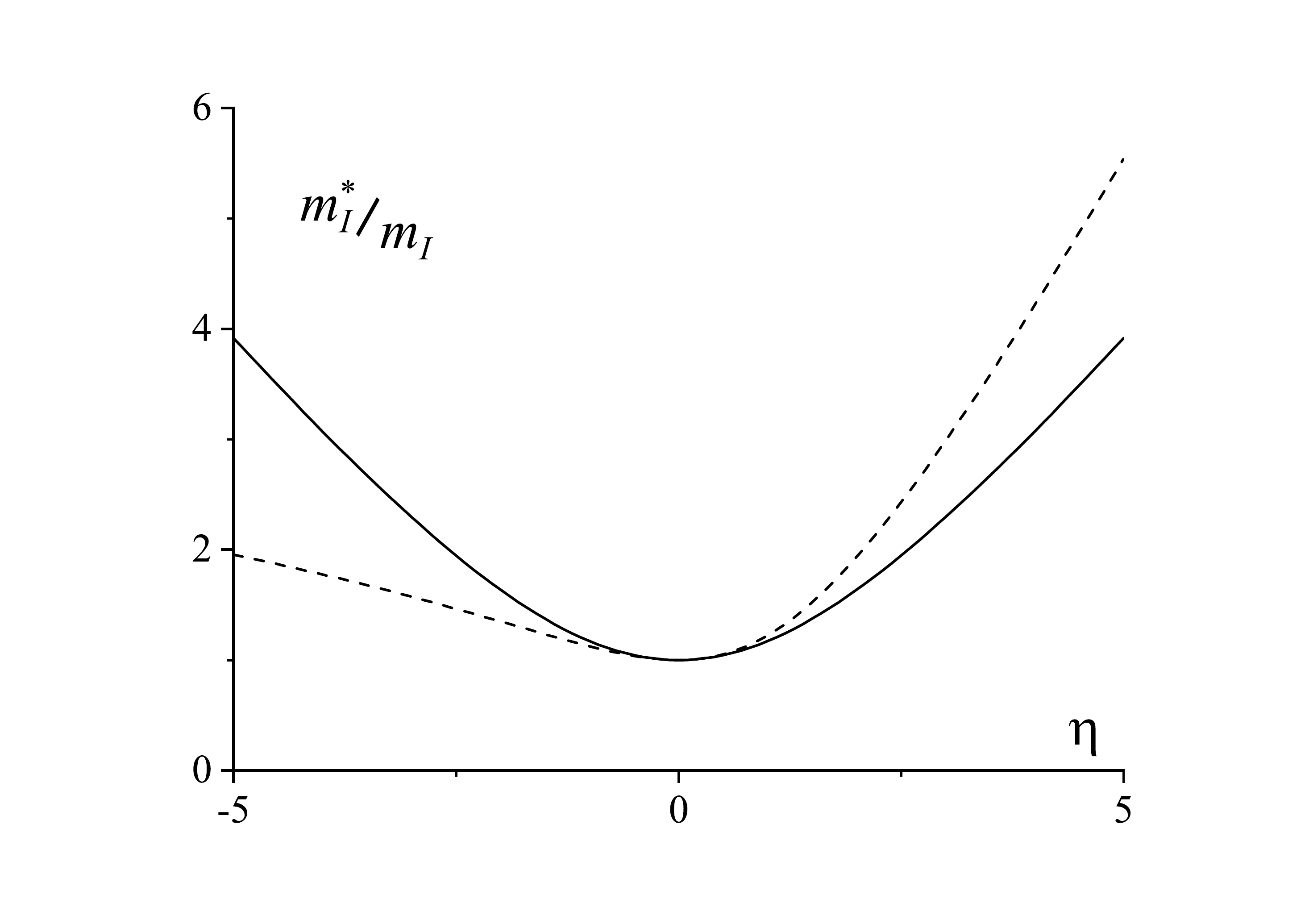}}
	\caption{Polaron effective mass in the Bose system with two- (solid line) and three-body (dashed line) interaction. Other parameters are the same as in Fig.~1.}
\end{figure}
\begin{figure}[h!]
	\centerline{\includegraphics
		[width=0.45\textwidth,clip,angle=-0]{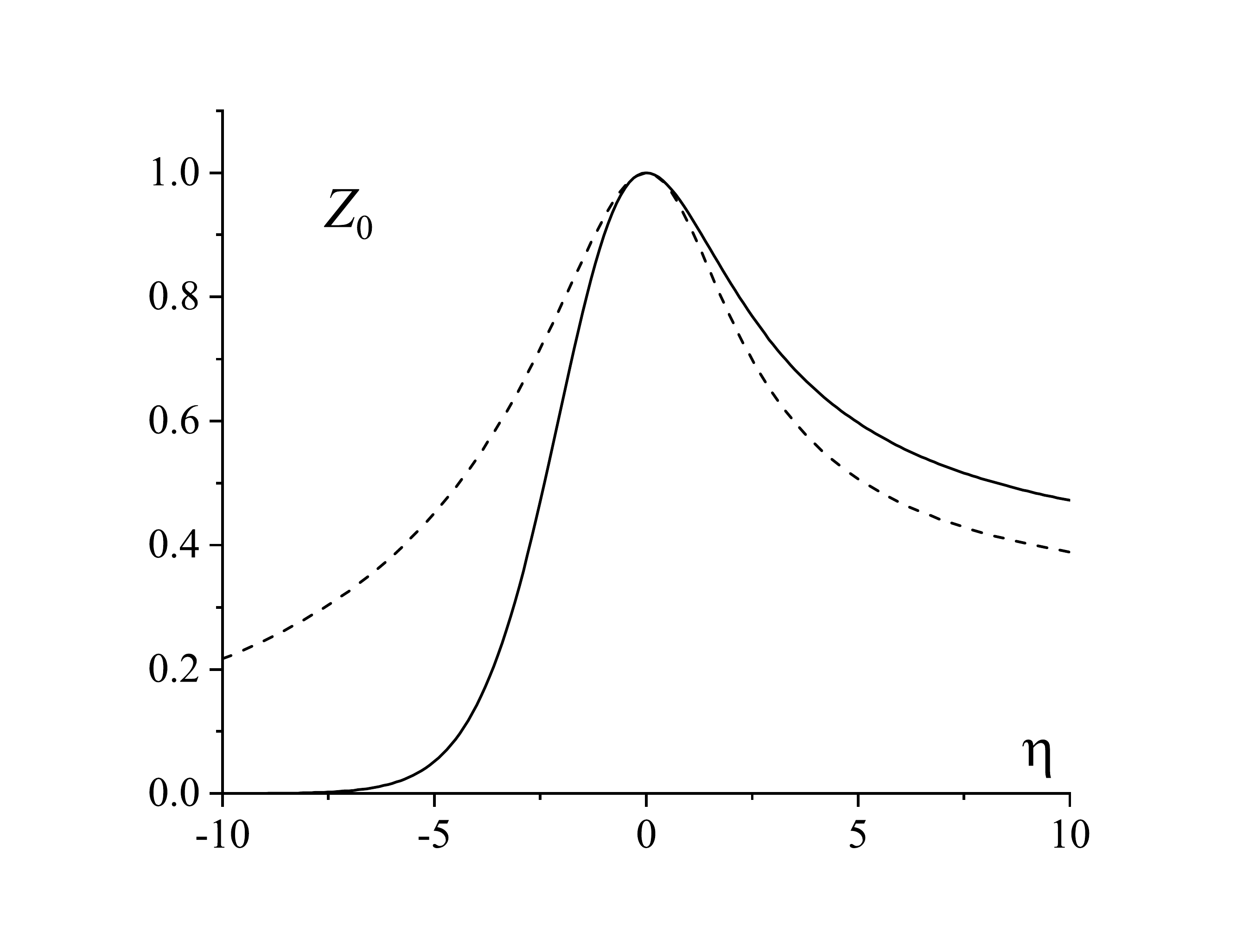}}
	\caption{Wave function renormalization $Z_0$ in the presence of motionless impurity $w=1$, $\eta_2=1$ (solid line) and $\eta_3=1$ (dashed line).}
\end{figure}
It is readily seen from Fig.~1 that the binding energy of the repulsive Bose polaron in the Bose gas with only two- or three-body interactions behaves qualitatively the same. The main differences appear for the attractive Bose polaron, especially at large $|g|$. Particularly the leading-order asymptote of the impurity energy changes from $\varepsilon_0\propto -|g|^3$ in the case of fully two-body repulsion between bosons  to $\varepsilon_0\propto -|g|^2$ in the case when only the tree-body interaction is present. This asymptotic behavoir of the binding energy survives in the general case, when both $g_2$ and $g_3$ are non-zero. The physics in this limit, however, is quite clear: the impurity forms a two-body bound state with particles of Bose medium and consequently the binding energy reaches $\varepsilon_0\to -m_rg^2/(2\hbar^2)$. This contradiction is not actually unexpected because the structure of the mean-field ansatz does not allow for the bound-state formation in the thermodynamic limit. 

The calculated effective mass (Fig.~2) reveals the differences between the two Bose environments more clearly. In the presence of the two-body repulsion ($g_3=0$) only, the quantity $\Delta_0$ determining the polaron effective mass is a very simple symmetric function of $g$ and is given by
\begin{eqnarray}\label{Delta_0}
\Delta_0|_{g_3= 0}=\frac{4m_r\bar{n}}{m_I\kappa}\left[\sqrt{1+\left(\kappa g/2\mu\right)^2}-1\right].
\end{eqnarray}
In the opposite limit ($g_2=0$), the effective mass differs for the repulsive and attractive Bose polarons (a cumbersome explicit analytic expression determining $\Delta_0$ can be obtained even in this case). It diverges when $g\to \pm \infty$, which is expected to be a qualitatively correct behavior for the repulsive boson-impurity interaction and totally incorrect for the attractive Bose polaron.

The preliminary conclusions drawn above for the effective mass can be easily extended to the quasiparticle residue (see Fig.~3), which was shown to demonstrate distinct dependence on the boson-impurity coupling parameter for the Bose gas with two- and three-body interaction, especially when $g<0$.

In order to reveal the interplay between two- and three-body forces in the one-dimensional Bose condensate on the behavior of a single impurity we build the graphical dependences of parameters of the polaron low-energy spectrum in Figs.~4-5.
\begin{figure}[h!]
	\centerline{\includegraphics
		[width=0.45\textwidth,clip,angle=-0]{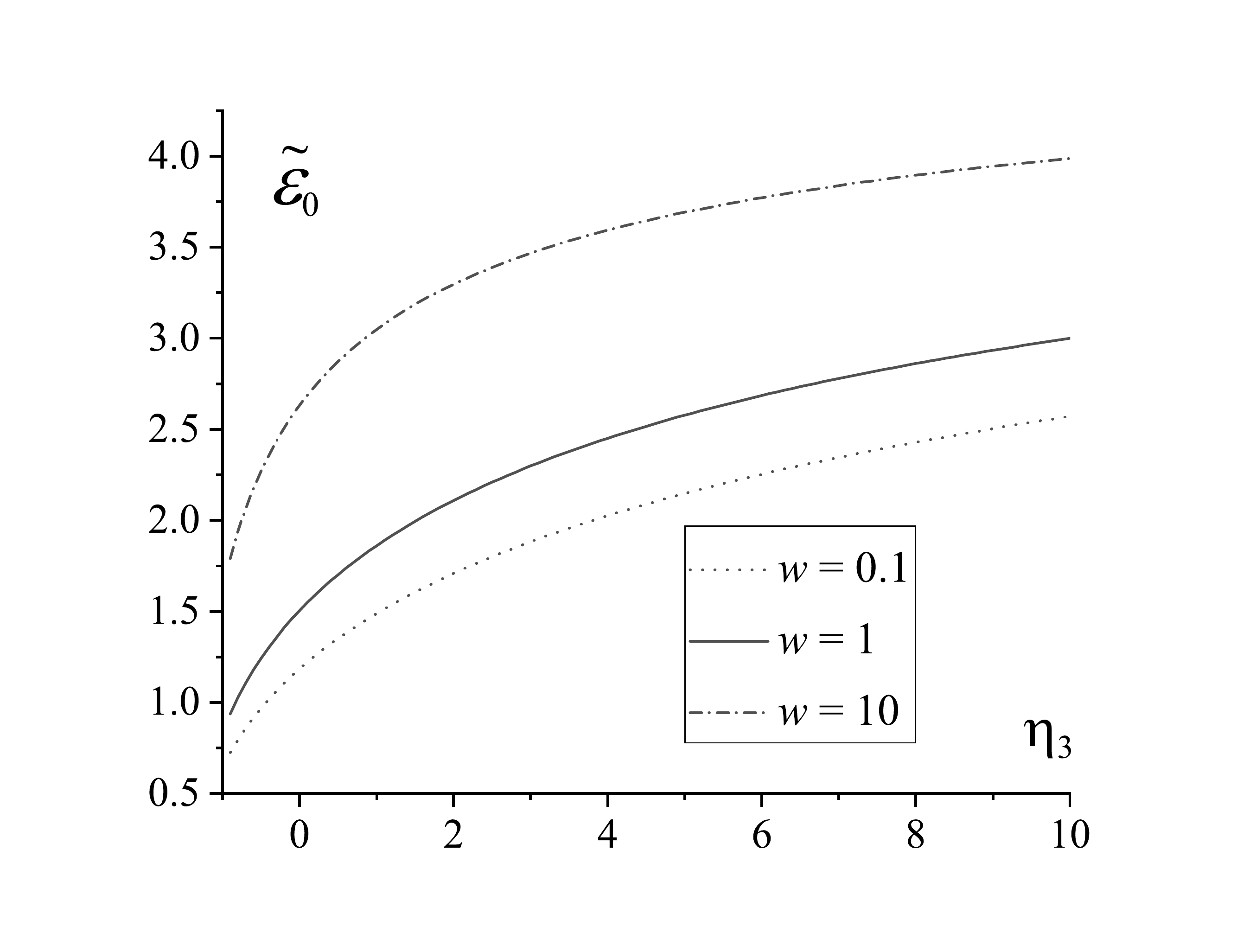}}
	\caption{Typical dependence of the polaron binding energy $\varepsilon_0$ (in units of $\hbar^2\bar{n}^2/m$) on the dimensionless three-body coupling constant $\eta_3$ presented here for three different mass ratios and fixed two-body boson-boson $\eta_2=1$ and boson-impurity $\eta=5$ couplings.}
\end{figure}
\begin{figure}[h!]
	\centerline{\includegraphics
		[width=0.45\textwidth,clip,angle=-0]{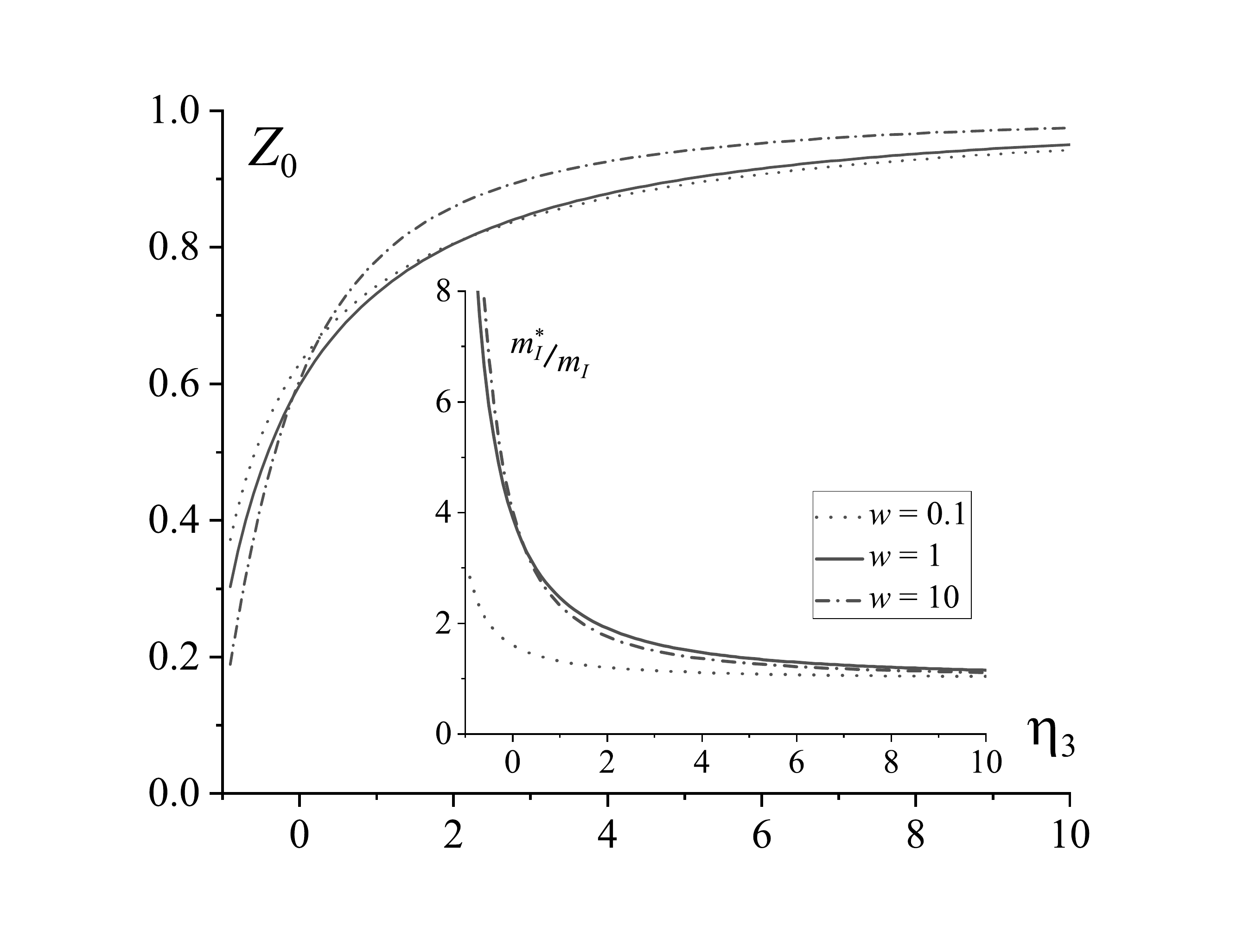}}
	\caption{Quasiparticle residue and impurity effective mass (inset) as functions of the three-body interaction parameter.}
\end{figure}
Particularly we fixed the strength of a boson-boson two-body coupling $\eta_2=1$ and focused on the case of a repulsive Bose polaron ($\eta=5$), where our mean-field consideration is expected to work correctly. Varying the three-body coupling parameter in the region, where the Bose system without impurity is stable $\eta_3\in (-1, \infty)$, at three different mass ratios $w=1/10;\, 1;\, 10$, we have numerically examined the impact of the three-body interaction (from weak attraction to strong repulsion) on properties of the system. Figures~4-5 outline the general tendency: an increase of the three-body repulsion leads to exhausting of the interaction effects. In particular, the impurity binding energy grows in the initial region and then reaches the constant value at large $g_3$, while the quasiparticle residue and the effective mass (see Fig.~5) monotonically tend to unity. Qualitatively the same behavior is observed while changing strength of the boson-impurity interaction (importantly that it should be the repulsive one).

It is also instructive to find out the influence of various mass ratios from static $w=0$ to light $w\to \infty$ impurity on the properties of one-dimensional Bose polaron. For these purposes we have plotted in Fig.~6
\begin{figure}[h!]
	\centerline{\includegraphics
		[width=0.45\textwidth,clip,angle=-0]{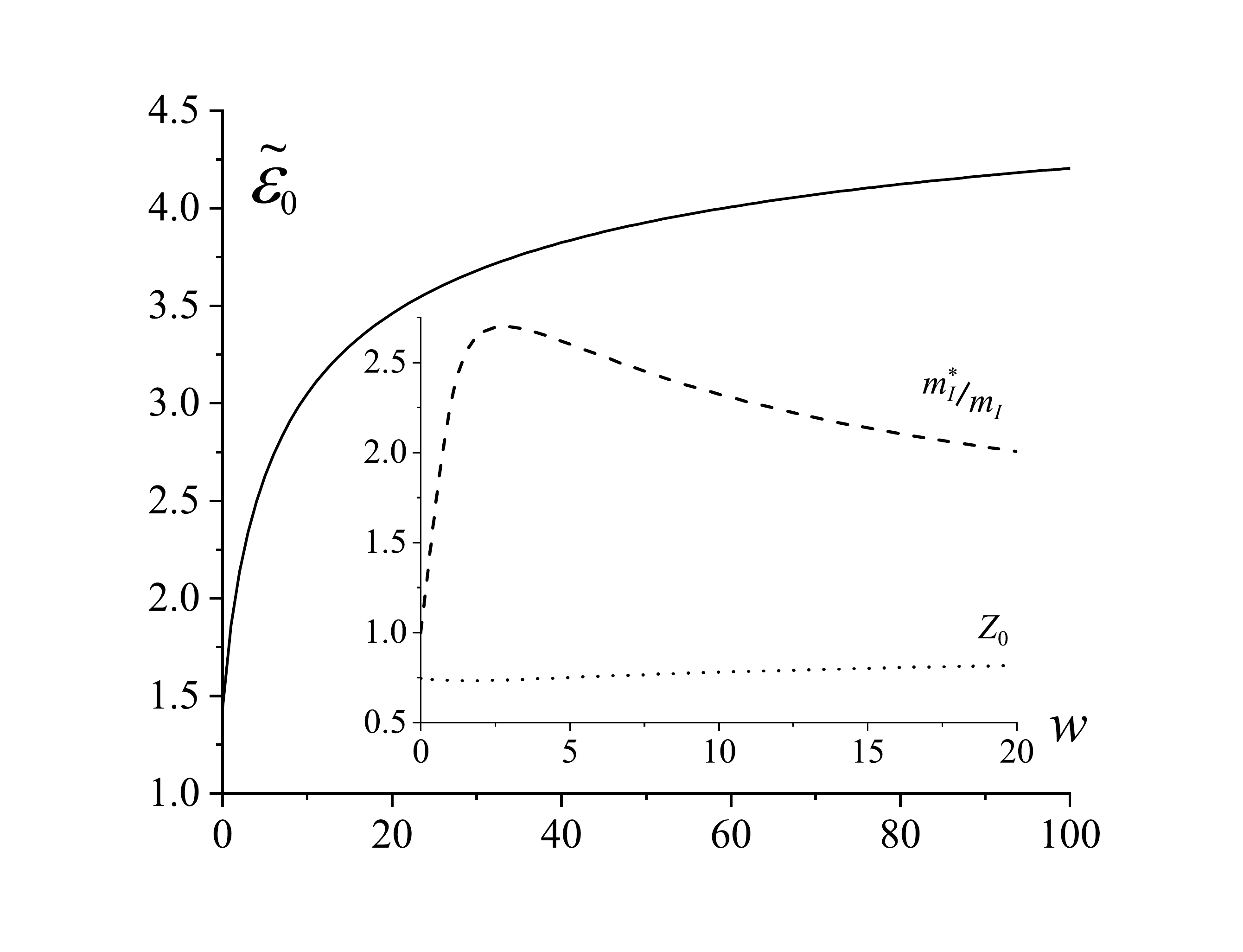}}
	\caption{Parameters of the low-energy Bose polaron spectrum ($\eta_2=\eta_3=1$ and $\eta=5$) versus the mass ratio parameter $w$.}
\end{figure}
the dependences of low-momentum parameters of the impurity spectrum on parameter $w$. In order to take into account both the two-body and three-body interaction effects on equal footing the dimensionless coupling constants were chosen as follows $\eta_2=\eta_3=1$. The strength of the boson-impurity short-range potential is assumed to be somewhat stronger $\eta=5$. As it is seen from the presented graphical dependenciesm the binding energy slowly grows with the increase of the mass ratio $w$, while the effective mass and the wave-function overlap tend to unity.

\section{Conclusions}
In conclusion, by using a simple and efficient method we exactly analyzed the mean-field behavior of an impurity immersed in the one-dimensional Bose gas with two- and three-body interaction between particles. We focused only on the low-momentum region of the polaron properties, namely, biding energy and quasiparticle residue and obtained a simple and obvious expression for the impurity effective mass in the surrounding Bose medium. On the mean-field level, all these observables are explicitly related to the density profile of bath particles that uniform distribution is disturbed by the presence of a point-like impurity. Remarkably, that this spatial distribution of the bath-forming bosons with contact two- and three-body interactions was found in our article to be given by a simple analytical formula. The obtained results may be summarized in the following statements: first, by allowing the three-boson interaction we totally change properties of the attractive Bose polaron, especially in the region of a strong boson-impurity attraction; second, even for the repulsive boson-impurity potential, where actually the adopted mean-field approach is qualitatively correct, the presence of three-body forces leads to essential exhausting of the interaction effects in the behavior of a mobile impurity. Finally, it should be noted that the mean-field construction of the 1D Bose polaron presented in this paper can be straightforwardly extended on higher dimensions and after some modifications on the description of 1D bipolaronic systems \cite{Dehkharghani_18}.

\begin{center}
	{\bf Acknowledgements}
\end{center}
We thank Prof.~Andrij~Rovenchak for useful comments. This work was partly supported by Project FF-30F (No.~0116U001539) from the Ministry of Education and Science of Ukraine.


\begin{thebibliography}{99}


\bibitem{Grusdt_Demler} F.~Grusdt, E.~Demler, arXiv preprint arXiv:1510.04934.


\bibitem{Schmid_et_al} S.~Schmid, A.~H\"arter and J.~H.~Denschlag, Phys.~Rev.~Lett. {\bf 105}, 133202 (2010).
\bibitem{Spethmann_et_al} N.~Spethmann, F.~Kindermann, S.~John, C.~Weber, D.~Meschede and A.~Widera, Phys.~Rev.~Lett. {\bf 109}, 235301 (2012).
\bibitem{Hu} M.~G.~Hu, ~M.~J.~Van~de~Graaff, D.~Kedar, J.~P.~Corson, E.~A.~Cornell, and D.~S.~Jin, Phys.~Rev.~Lett. {\bf 117}, 055301 (2016).
\bibitem{Jorgenzen} N.~B.~J\o{}rgensen, L.~Wacker, K.~T.~Skalmstang, M.~M.~Parish, J.~Levinsen, R.~S.~Christensen, G.~M.~Bruun and J.~J.~Arlt, Phys.~Rev.~Lett. {\bf 117} 055302 (2016).	



\bibitem{Rath} S.~P.~Rath and R.~Schmidt, Phys.~Rev.~A {\bf 88}, 053632 (2013).
\bibitem{Shashi} A.~Shashi, F.~Grusdt, D.~A. Abanin, and E.~Demler,
Phys.~Rev.~A {\bf 89}, 053617 (2014).
\bibitem{Li} W.~Li and S.~Das~Sarma, Phys.~Rev.~A {\bf 90}, 013618 (2014).
\bibitem{Christensen} R.~S.~Christensen, J.~Levinsen, and G.~M.~Bruun, Phys.~Rev.~Lett. {\bf 115}, 160401 (2015).
\bibitem{Grusdt_15} F.~Grusdt, Y.~E.~Shchadilova, A.~N.~Rubtsov and
E.~Demler, Sci.~Rep. {\bf 5}, 12124 (2015).
\bibitem{Volosniev_15} A.~G.~Volosniev, H.-W.~Hammer, and N.~T.~Zinner,
Phys.~Rev.~A 92, 023623 (2015).
\bibitem{Vlietinck_et_al} J. Vlietinck W.~Casteels, K.~Van~Houcke, J.~Tempere, J.~Ryckebusch and J.~T.~Devreese, New~J.~Phys. {\bf 17},  033023 (2015).
\bibitem{Ardila_1} L.~A.~Pe\~na~Ardila and S.~Giorgini, Phys.~Rev.~A {\bf 92}, 033612 (2015).
\bibitem{Shchadilova} Y.~E.~Shchadilova, F.~Grusdt, A.~N.~Rubtsov, and E.~Demler
Phys.~Rev.~A {\bf 93}, 043606 (2016).
\bibitem{Ardila_2} L.~A.~Pe\~na~Ardila and S.~Giorgini, Phys.~Rev.~A {\bf 94}, 063640 (2016).
\bibitem{Kain_Ling_16} B.~Kain and H.~Y.~Ling, Phys.~Rev.~A {\bf 94}, 013621 (2016).
\bibitem{Grusdt_17} F.~Grusdt, R.~Schmidt, Y.~E.~Shchadilova, and E.~Demler,
Phys.~Rev.~A {\bf 96}, 013607 (2017).
\bibitem{Panochko_17} G.~Panochko, V.~Pastukhov, I.~Vakarchuk, Condens.~Matter~Phys. {\bf 20}, 13604 (2017).
\bibitem{Lampo} A.~Lampo, S.~H.~Lim, M.~\'A.~Garc\'ia-March, M.~Lewenstein, Quantum {1}, 30 (2018).
\bibitem{Grusdt_18} F.~Grusdt, K.~Seetharam, Y.~Shchadilova, and E.~Demler,
Phys.~Rev.~A {\bf 97}, 033612 (2018).
\bibitem{Pastukhov_CrBP} V.~Pastukhov, J.~Phys.~A: Math.~Theor. {\bf 51}, 195003 (2018).
\bibitem{Panochko_18} G.~Panochko, V.~Pastukhov, I.~Vakarchuk, Int.~J.~Mod.~Phys.~B {\bf 32}, 1850053 (2018).
\bibitem{Mehboudi} M.~Mehboudi, A.~Lampo, C.~Charalambous, L.~A.~Correa, M.~\'A.~Garc\'ia-March, M.~Lewenstein, arXiv preprint 	arXiv:1806.07198.

\bibitem{Lychkovskiy_18} O.~Lychkovskiy, O.~Gamayun, and V.~Cheianov, AIP
Conf. Proc. {\bf 1936}, 020024 (2018).

\bibitem{Burovski} E.~Burovski, V.~Cheianov, O.~Gamayun, and O.~Lychkovskiy
Phys.~Rev.~A {\bf 89}, 041601(R) (2014).
\bibitem{Gamayun} O.~Gamayun, O.~Lychkovskiy, and V.~Cheianov
Phys.~Rev.~E {\bf 90}, 032132 (2014).

\bibitem{Catani_et_al} J.~Catani, G.~Lamporesi, D.~Naik, M.~Gring, M.~Inguscio, F.~Minardi, A.~Kantian, and T.~Giamarchi, Phys.~Rev.~A {\bf 85}, 023623 (2012).

\bibitem{Parisi} L.~Parisi and S.~Giorgini, Phys.~Rev.~A {\bf 95}, 023619 (2017).
\bibitem{Grusdt} F.~Grusdt, G.~E.~Astrakharchik, E.~A.~Demler, New~J.~Phys. {\bf 19}, 103035 (2017).
\bibitem{Ovchinnikov} M.~Ovchinnikov and A.~Novikov, J.~Chem.~Phys. {\bf 132}, 214101 (2010).
\bibitem{Dehkharghani} A.~S.~Dehkharghani, A.~G.~Volosniev, and N.~T.~Zinner, Phys.~Rev.~A {\bf 92}, 031601(R) (2015).
\bibitem{Volosniev} A.~G.~Volosniev, H.-W.~Hammer, Phys.~Rev.~A {\bf 96},
031601(R) (2017).
\bibitem{Pastukhov_1D} V.~Pastukhov, Phys.~Rev.~A {\bf 96}, 043625 (2017).
\bibitem{Kain_Ling_18} B.~Kain and H.~Y.~Ling, Phys.~Rev.~A {\bf 98}, 033610 (2018).
\bibitem{Mistakidis} S.~I.~Mistakidis, A.~G.~Volosniev, N.~T.~Zinner, 
P.~Schmelcher, arXiv preprint arXiv:1809.01889.


\bibitem{McGuire_65} J.~B.~McGuire, J.~Math.~Phys. {\bf 6}, 432 (1965).
\bibitem{McGuire_66} J.~B.~McGuire, J.~Math.~Phys. {\bf 7}, 123 (1966).

\bibitem{Sekino} Y.~Sekino and Y.~Nishida, Phys.~Rev.~A {\bf 97}, 011602(R) (2018).
\bibitem{Pastukhov_3BI} V.~Pastukhov, arXiv preprint arXiv:1807.07106.
\bibitem{Pricoupenko} L.~Pricoupenko, Phys.~Rev.~A {\bf 97}, 061604(R) (2018).
\bibitem{Guijarro} G.~Guijarro, A.~Pricoupenko, G.~E.~Astrakharchik, J.~Boronat, and D.~S.~Petrov, Phys.~Rev.~A {\bf 97}, 061605(R) (2018).
\bibitem{Nishida} Y.~Nishida, Phys.~Rev.~A {\bf 97}, 061603(R) (2018).

\bibitem{LLP} T.~D.~Lee, F.~E.~Low, and D.~Pines, Phys. Rev. {\bf 90}, 297 (1953).

\bibitem{Astrakharchik_04} G.~E.~Astrakharchik and L.~P.~Pitaevskii, Phys.~Rev.~A
{\bf 70}, 013608 (2004).
\bibitem{Gross} E.~P.~Gross, J.~Math.~Phys. {\bf 4}, 195 (1963).
	
\bibitem{Pastukhov_2DBP} V.~Pastukhov, J.~Phys.~B:~At.~Mol.~Opt.~Phys. {\bf 51} 155203 (2018).	


\bibitem{Carr} L.~D.~Carr, C.~W.~Clark, and W.~P.~Reinhardt, Phys.~Rev.~A {\bf 62},
063610 (2000).
\bibitem{DAgosta} R.~D'Agosta, B.~A.~Malomed, and C.~Presilla, Phys.~Lett.~A {\bf 275}, 424 (2000).

\bibitem{Dehkharghani_18} A.~S.~Dehkharghani, A.~G.~Volosniev, and N.~T.~Zinner,
Phys.~Rev.~Lett. {\bf 121}, 080405 (2018).



\end{thebibliography}
\end{document}